\documentclass[twoside,11pt]{article}

\usepackage{cite}      
\usepackage{graphicx}  
\usepackage{psfrag}    
\usepackage{subfigure} 
\usepackage{url}       
\usepackage{amsmath}   
\usepackage{amsfonts}
\usepackage{epsfig}
\usepackage{amssymb,comment}
\usepackage{theorem}
\newtheorem{Theorem}{Theorem}
\newtheorem{Corollary}{Corollary}
\newtheorem{Lemma}{Lemma}
\newtheorem{Example}{Example}
\newtheorem{Proposition}{Proposition}

\newtheorem{Definition}{Definition}
\newenvironment{proof}{{\em Proof:} \ \ }{\begin{flushright}$\Box$\end{flushright}}

\newcommand{\Data}{{\mathcal{D}_n}}
\newcommand{\J}{\mathcal{J}}
\newcommand{\Lag}{\mathcal{L}}
\newcommand{\Hyp}{\Gamma}  
\newcommand{\Hypm}{{\mathcal{H}_1}}
\newcommand{\Hyps}{{\mathcal{H}_2}}
\newcommand{\R}{\mathbb{R}}
\newcommand{\X}{X}
\newcommand{\E}{U}
\newcommand{\Exp}{\mathbb{E}}
\newcommand{\Y}{Y}

\newcommand{\Tube}{\mathcal{T}_{m,s}}
\newcommand{\Risk}{\mathcal{R}}

\newcommand{\CH}{{\rm CH}}
\newcommand{\suchthat}{\mbox{\ s.t. \ }}

\DeclareMathOperator*{\argmin}{\arg\min}

\DeclareMathOperator*{\ind}{\Bbb{I}}

\begin{document}
%
\title{Support and Quantile Tubes}
\author{Kristiaan Pelckmans,
  Jos De Brabanter,
  Johan A.K. ~Suykens,
  and~Bart~De~Moor.
\thanks{Pelckmans {\em et al.} are with 
  KULeuven-ESAT-SCD/sista, Kasteelpark Arenberg 10, 
  Leuven - B-3001, Belgium}}
%

\maketitle

\begin{abstract}
  This correspondence studies an estimator of the conditional 
  support of a distribution underlying a set of i.i.d. observations. 
  The relation with mutual information is shown via an extension of 
  Fano's theorem in combination with a generalization bound
  based on a compression argument. 
  Extensions to estimating the conditional quantile interval, and 
  statistical guarantees on the minimal convex hull are given.

  {\bf Keywords}: - 
  Statistical Learning,
  Fano's inequality,
  Mutual Information,
  Support Vector Machines
\end{abstract}

\section{Introduction}

Given a set of paired observations 
$\Data=\{(X_i,Y_i)\}_{i=1}^n\subset\R^d\times\R$ 
which are i.i.d. copies of a random vector $(\X,\Y)$
possessing a fixed but unknown joint distribution $F_{\X\Y}$, 
this letter concerns the question which values the random variable $Y$ can 
possibly/likely take given a covariate $\X$.
This investigation on predictive tolerance intervals 
is motivated as one is often interested in
other characteristics of the joint distribution than the conditional expectation (regression):
e.g. in econometrics one is often more interested in the volatility
of a market than in its precise prediction. In environmental sciences 
one is typically concerned with the extremal behavior 
(i.e. the min or max value) of a magnitude, and its respective conditioning 
on related environmental variables.

The main contribution of this letter is the extension to Fano's classical inequality
(see e.g. \cite{cover91}, p. 38)
which gives a lower-bound to the mutual information of two random variables.
This classical result is extended towards a setting of learning theory
where random variables have an arbitrary fixed distribution. 
The derivation yields a non-parametric estimator of the mutual information possessing 
a probabilistic guarantee which is derived using a classical compression argument.
The described relationship differs from other results relating 
estimators and mutual information as e.g. using Fisher's information matrix \cite{cover91}
or based on Gaussian assumptions as e.g. in \cite{guo05},
as a distribution free context is adopted.  
As an aside, 
(i) an estimator of the conditional support is derived 
and is extended to the setting of conditional quantiles, 
(ii) its theoretical properties are derived, 
(iii) the relation to the method of the minimal convex hull is made explicit,
and (iv) it is shown how the estimate can be computed efficiently by 
solving a linear program.

While studied in the literature e.g. on quantile regression 
\cite{koenker05}, we argue that this question can be approached  
naturally from a setting of statistical learning theory, pattern recognition
and Support Vector Machines (SVM), see \cite{vapnik98,devroye96} for an overview.
A main conceptual difference with the existing literature on classical regression and 
other predictor methods is that no attempt is made whatsoever to reveal an underlying 
conditional mean (as in regression), conditional quantile (as in quantile regression),
or minimal risk point prediction of the dependent variable (as in pattern recognition). 
Here we target instead (the change of) the rough contour of the conditional distribution. 
This implies that one becomes interested in 
(i) to what extent the estimated conditional support of the tube is conservative 
(i.e. does it overestimate the actual conditional support?), and 
(ii) what is the probability of covering the actual conditional support
(i.e. to what probability a new sample can occur outside the estimated interval).

\begin{figure}
  \begin{center}
    \scalebox{.5}{\psfig{figure=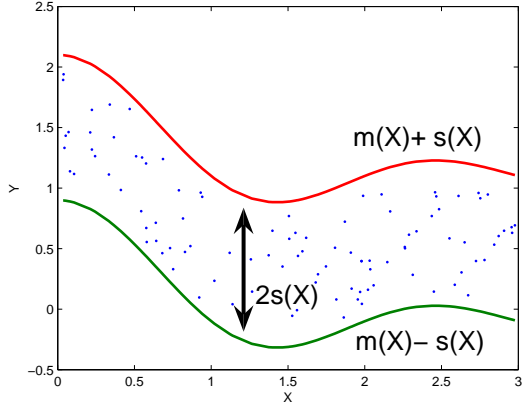}}
  \end{center}
  \caption{\sl\small 
    Example of a support vector tube based on a finite sample of a 
    bivariate random variable $(X,Y)$.
    A tube $\Tube$ is defined as the conditional interval 
    $\Tube(X) = [m(X)-s(X), \ \ m(X)+s(X)]$ with 
    width $2s(x)$.
  }
  \label{fig.ex}
\end{figure}

Section II proofs the main result, and explores the relation with the convex hull.
From a practical perspective, Section III provides further insight in 
how the optimal estimate can be found efficiently by solving a linear program.

\section{Support and Quantile Tubes}

\subsection{Support Tubes and Risk}


\begin{Definition}[Support and Quantile Tubes]
  Given a set of data $\Data$ which are sampled i.i.d. from a fixed but unknown 
  joint distribution $F_{\X\Y}$.
  Let $\Hypm\subset\{m:\R^d\rightarrow\R\}$ 
  and $\Hyps\subset\{s:\R^d\rightarrow\R^+\}$ be proper function spaces where the latter is 
  restricted to positive functions and $\Hyps\subset\Hypm$. 
  Let $p(\R)$ be the powerset of $\R$ such that $p(\R) = \{V\subset\R\}$. 
  The class of tubes $\Hyp(\Hypm,\Hyps)$ is defined as 
  \begin{multline}
    \Hyp(\Hypm,\Hyps) = 
    \left\{\Tube:\R^d\rightarrow p(\R), \ m\in\Hypm, s\in\Hyps\ \ \Big| \ \right.\\\left. 
      \Tube(x) = [m(x)-s(x), \ m(x)+s(x)] \right\} 
      \label{eq.st}
  \end{multline}
  abbreviated as $\Tube=m\pm s$. 
  A tube $\Tube\in\Hyp(\Hypm,\Hyps)$ 
  is a true support tube (ST) of a joint distribution $F_{\X\Y}$ if 
  the equality $P(\Y\in\Tube(\X)) = 1$ holds. 
  Similarly a tube $\Tube\in\Hyp(\Hypm,\Hyps)$ 
  is a true quantile tube (QT) for $F_{\X\Y}$ 
  of level $0<\alpha<1$
  if $P(\Y\in\Tube(\X)) \geq 1-\alpha$.
\end{Definition}
Let the indicator $\ind(\Y\not\in\Tube(\X))$ be equal to one if $\Y\not\in\Tube(\X)$ 
and zero otherwise. We define the risk of a candidate ST for given joint 
distribution as follows
\begin{equation}
  \Risk(\Tube; F_{\X\Y}) 
  = E\left[ \ind\left( Y\not\in \Tube(X)\right)\right] 
  = P\left( Y\not\in \Tube(X)\right), 
  \label{eq.risk}
\end{equation}
where the expectation is taken over the random variables $X$ and $Y$ 
with joint distribution $F_{XY}$.
Its empirical counterpart becomes
$\Risk_n(\Tube;\Data) = \frac{1}{n} \sum_{i=1}^n \ind\left(Y_i\not\in \Tube(X_i)\right)$.
The study of support tubes based on empirical samples will yield bounds of the form 
\begin{equation}
  P \left( \sup_{\Tube\in\Hyp} \Risk(\Tube;F_{\X\Y}) \geq  \epsilon \right) 
  \leq \eta(\epsilon;\Hyp(\Hypm,\Hyps)),
  \label{eq.bound}
\end{equation}
where $0<1-\epsilon<1$ is the {\em probability of covering}
the tube and where the function $\eta(\cdot;\Hyp(\Hypm,\Hyps)): [0,1]\rightarrow [0,1)$
expresses the {\em confidence level} in the probability of covering.

\subsection{Generalization Bound}

For now, we focus on the case of the ST, extensions specific to the 
QT are described in the next subsection.
Assume a given hypothesis class $\Hyp(\Hypm,\Hyps)$ 
of STs. Consider an algorithm constructing 
a ST - say $\Tube$ - with zero empirical risk $\Risk_n(\Tube;\Data)=0$.
The generalization performance can be bounded using a geometrical argument
which was also used for deriving the compression bound outlined in 
\cite{littlestone86},\cite{floyd95},
and refined in various publications as e.g. \cite{luxburg04}.

\begin{Theorem}[Compression Bound on Risk of a ST]
  Let $\Data$ be i.i.d. sampled from a fixed but unknown joint distribution $F_{XY}$.
  Consider the class of tubes $\Hyp$ where each tube 
  $\Tube$ is uniquely determined by $D$ appropriate samples 
  (i.e., $\Tube$ can be {\em 'compressed'} to $D$ samples). 
  Let $n_D=n-D$ denote the number of remaining samples.
  Then, with probability exceeding $1-\delta<1$, 
  the following inequality holds for any $\Tube$ where 
  $\Risk_n(\Tube;\Data)=0$:
  \begin{equation}
    \sup_{\Risk_n(\Tube;\Data)=0} 
    \Risk(\Tube; F_{XY}) \\ \leq 
    \frac{\log\left(K_{n,D}(\Hyp)\right) + \log\left(\frac{1}{\delta}\right)}{n-D}
    \triangleq \epsilon(\delta,D,n),
    \label{eq.tolerance}
  \end{equation}
  where we define $K_{n,D}(\Hyp)$ as 
  \begin{equation}
    K_{n,D}(\Hyp) = \binom{n}{D} (2^{D-1}-1) \leq \left(\frac{2 n e}{D}\right)^D.
    \label{eq.KnD}
  \end{equation}
\end{Theorem}
\begin{proof}
  At first, fix a ST determined by $D$ samples - say the first $D$ samples 
  $\{(X_1,Y_1),\dots,(X_D,Y_D)\}$ - denoted as $\Tube^D$.
  Assume $F_{XY}$ is such that the actual risk of 
  this tube is larger than a given value $0<\epsilon<1$ 
  such that $\Risk(\Tube^D;F_{XY}) \geq \epsilon$. 
  Then the chance that the remaining $n-D$
  i.i.d. samples $\{(X_{D+1}Y_{D+1}),\dots,(X_n,Y_n)\}$ are 
  by chance consistent with $\Tube^D$, is lower than
  $\prod_{i=D+1}^n  P\left(Y_i\in\Tube^D(X_i)\right) \leq (1-\epsilon)^{n-D}$.
  This can be bounded as follows 
  \begin{equation}
    P\left(\Risk(\Tube^D;F_{XY})\geq \epsilon\right) 
    \leq (1-\epsilon)^{n-D} 
    \leq e^{-(n-D)\epsilon},
    \label{eq.chance}
  \end{equation}  
  making use of the classical binomial bound, see e.g. \cite{devroye96}.
  The finite number of tubes which can be compressed without loss of information to $D$ points
  can be bounded using a geometrical argument. 
  Given $D$ points, every point can be used 
  to interpolate either the upper-function $m+s$, or the lower-function $m-s$.
  However, switching the assignments of all points simultaneously leads to the same ST, 
  and the case of all points assigned to the same (upper- or lower-) function 
  does not result in a unique tube neither.
  Therefor, the number of ST which can be determined using $D$ samples out of 
  $n$ - denoted as $K_{n,D}(\Hyp)$ - can be bounded as follows:
  \begin{eqnarray}
    K_{n,D}(\Hyp)
    &\leq& \binom{n}{D} (2^{D-1}-1) \notag \\
    &\leq&  \left(\frac{ne}{D}\right)^D (2^{D-1}-1)
    \leq \left(\frac{2 n e}{D}\right)^D
    \label{eq.KnD2}
  \end{eqnarray}
  where the inequality $\binom{n}{D}\leq (\frac{ne}{D})^D$
  of the binomial coefficient is used.
  Combining  (\ref{eq.chance}) and (\ref{eq.KnD}),
  and inverting the statement as classical proofs the result.
\end{proof}

A crucial element for this result is that it is known a priori that such a 
tube with zero empirical risk exists independently from the data at hand
(realizable case), this assumption is fulfilled by construction. 
Although combinatorial in nature (any found hypothesis $\Hyp$ should
be determined entirely by a subset of $D$ chosen examples), 
it is shown in the next section how this property holds for a simple estimator 
which can be estimated efficiently as a standard linear program.

\begin{Example}[Tolerance level]
The following example indicates the practical use of this result:
given $n=200$ i.i.d. samples 
with a corresponding class of hypotheses each 
determined by three samples ($D=3$ and thus $K_{n,D}(\Hyp)\leq 3*10^8$). 
Fixing the tolerance level as  $\delta = 95\%$, 
one can state that the true risk will not be higher than $0.1049$.
This result can be used in practice as follows.
Given an observed set of i.i.d. samples $\Data = \{(X_i,Y_i)\}_{i=1}^{200}\subset\R\times\R$,
compute the tube $\widehat\Tube = \hat{w} x\pm \hat{t}$ with $\hat{t}>0$, $w\in\R$ 
and $\Risk_n(\widehat\Tube;\Data)=0$.
When a new sample $X_j\in\R$ arrives, 
then predict that the corresponding $Y_j\in\R$ will lie in the interval $\hat{w}X_j \pm \hat{t}$.
Then we are reasonably sure (with a probability of $0.95$) that this assertion will 
hold in at least $89.51\%$ of the cases when the number $n_v$ of samples 
of data $\{X_j\}_{j=1}^{n_v}$ goes to infinity.
\end{Example}

A similar result can be obtained using the classical theory of 
non-parametric tolerance intervals, 
as initiated in \cite{wilks41}, see e.g. \cite{rice88}.
\begin{Corollary}[Bound by Order Statistics]
  Let $\Data$ be i.i.d. samples from a fixed but unknown joint distribution $F_{XY}$.
  Consider the class of tubes $\Hyp$ where each tube 
  $\Tube$ is uniquely determined by $D$ appropriate samples. 
  Then, with probability higher than $1-\delta<1$, 
  the following inequality holds for any $\Tube$ where 
  $\Risk_n(\Tube;\Data)=0$:  
  \begin{multline}
    P\left( 
      \sup_{\Risk_n(\Tube;\Data)=0} \Risk(\Tube; F_{XY}) \geq \epsilon \right) \\
    \leq 
    K_{n,D}(\Hyp)\left(n(1-\epsilon)^{n-1} - (n-1)(1-\epsilon)^n \right),
  \end{multline}
  where $K_{n,D}(\Hyp)$ is defined as in Theorem 1.
\end{Corollary}
\begin{proof}
  Consider at first a fixed tube $\Tube^\ast$.
  After projecting all samples $\{(X_i, Y_i)\}_{i=1}^n$
  to the univariate sample $R_i = m(X_i) - Y_i$,
  it is clear that a minimal tube with fixed $m$ will 
  have borders $\min(R_i)$ and $\max(R_i)$.
  Note that now $P(R\not\in[\min(R_i),\max(R_i)])$ equals $\Risk(\Tube^\ast; F_{XY})$.
  Application of the standard results as in \cite{wilks41}
  for such tolerance intervals gives 
  \begin{equation}
    P\Big( P(R\not\in[\min(R_i),\max(R_i)]) \geq \epsilon\big) \\
    \leq n(1-\epsilon)^{n-1}- (n-1)(1-\epsilon)^n
    \label{eq.tolerance}
  \end{equation}
  Application of the union bound over all hypothesis $\Hyp$ 
  as in (\ref{eq.KnD}) gives the result.
\end{proof}
Remark that this bound is qualitatively very similar to the previous one.
As a most interesting aside, 
the previous result implies a generalization bound on the minimal convex hull,
i.e. a bound on the probability mass contained in the minimal Convex Hull (CH) 
of an i.i.d. sample. We consider the planar case, 
the extension to higher dimensional case follows straightforwardly.
Formally, one may define the minimal planar convex hull ${\rm CH}(\Data)$
of a sample $\Data = \{(X_i,Y_i)\}_{i=1}^n$
as the minimal subset of $\R\times\R$ containing all samples $(X_i,Y_i)\in\R\times\R$, 
and all convex combinations of any set of samples. 

\begin{Theorem}[Probability Mass of the Planar Convex Hull]
  Let $\Data$ contain i.i.d. samples of a random variable 
  $(\X,\Y)\subset\R\times\R$.
  Then with probability exceeding $1-\delta<1$, 
  the probability mass outside the minimal convex 
  hull ${\rm CH}(\Data)$ is bounded as follows
    \begin{equation}
      P\left( (X,Y)\not\in{\rm CH}(\Data) \right) 
      \leq 
      \frac{3 \log(n) -1.5122 - \log(\delta)}{n-3}.
      \label{eq.cv.bound}
    \end{equation}
\end{Theorem}
\begin{proof}
  The key element of the proof is found in the fact that the 
  CH is the intersection of all linear support tubes in 
  $\Hyp$ with minimal (constant) width having zero empirical risk.
  Let $\#\CH(\Data)$ denote this intersection, formally,
  \begin{equation}
    (X,Y) \in \#\CH(\Data) \\
    \Leftrightarrow 
    Y \in \Tube(X), \ \ \forall \Tube: \ \ \Risk_n(\Tube;\Data)=0.
    \label{eq.chvstube}
  \end{equation}
  Now we proof that $\#\CH(\Data) = \CH(\Data)$.
  Assume at first that $\#\CH(\Data) \subset \CH(\Data)$,
  then a point $(X,Y)\in\CH(\Data)$ exists where $(X,Y)\not\in\#\CH(\Data)$,
  but this is in contradiction to the assertion that $\CH(\Data)$ 
  should be minimal: indeed also $\#\CH(\Data)$ is convex 
  (an intersection of convex sets), and contains all samples
  by construction.

  Conversely, assume that $\CH(\Data) \subset \#\CH(\Data)$,
  then a point $(X,Y)\in\#\CH(\Data)$ exist where $(X,Y)\not\in\CH(\Data)$,
  and the point $(X,Y)$ is included in all tubes $\Tube$
  having $\Risk_n(\Tube;\Data)=0$. 
  By definition of the convex hull $(X,Y)\not\in\Data$, neither can it be a 
  convex combination of any set of samples.
  Now, by the supporting hyperplane theorem (see e.g. \cite{rockafeller70}), there 
  exists a linear hyperplane separating this point 
  from the minimal convex hull. Constructing a tube $\Tube$ where 
  $m+s$ equals this supporting plane, and with width large enough such 
  that $\Risk_n(\Tube;\Data) = 0$ contradicts the assumption, proving the result. 

  Now, note that by definition the following inequality holds
  \begin{equation}
    P\Big( (X,Y)\not\in\#\CH(\Data) \Big) 
    = \sup_{\Risk_n(\Tube;\Data)=0} \Risk(\Tube; F_{XY}).
  \end{equation}
  Moreover, the set of linear tubes in $\R^2$ with fixed 
  width can be characterized by a set containing exactly $D=3$ samples 
  as proven in the following section.
  Finally, specializing the result of Theorem 1 in (\ref{eq.tolerance}) gives the result.
\end{proof}

Note that classically the expected probability mass of a CH is 
expressed in terms of the expected number of extremal points of the data cloud
\cite{efron65}. Interestingly, the literature on statistical learning
studies the number of extreme points in estimators as 
an (empirical) measure of complexity of an hypothesis space,
note e.g. the correspondence between Theorem 12 in \cite{vapnik98}
and Theorem 2 in \cite{efron65}, and  the coding interpretation of SVMs, 
see e.g. \cite{floyd95,vapnik98,luxburg04}.
A disadvantage of the mentioned approach appears that the expected number of extremal 
points of the convex hull is a quantity which is difficult 
to characterize a priori (without seeing the data), 
without presuming restrictions on the underlying distribution \cite{devroye96}.
The key observation of the previous theorem is that this number can be bounded 
by decomposing the minimal convex hull as the intersection of a set of linear tubes.

\subsection{Support Tubes and Mutual Information}

At first, a technical Lemma is proven which will play a major role 
in the main result of the paper stated below.
\begin{Lemma}[Upper-bound to the Conditional Entropy]
  Let $\Tube:\R^d \rightarrow V \subset\R$ 
  be a fixed tube, then one has
  \begin{equation}
    H(Y| (X,Y)\in\Tube(X) ) \leq \Exp[\log(2s(X))].
    \label{eq.ch}
  \end{equation}
\end{Lemma}
\begin{proof}
  The proof follows from the following inequality,
  for a fixed $x\in\R^d$ it holds that
  \begin{equation}
    H(Y|Y\in\Tube(x)) \leq \log(2s(x))
    \label{eq.ch1}
  \end{equation}
  following the fact that the uniform distribution has 
  maximal entropy over all distributions in a fixed interval.
  The conditional distribution is then defined as follows
  \begin{multline}
    H(Y|(X,Y)\in\Tube(X)) 
    = \int H(Y|X=x,Y\in\Tube(x)) \ dF_X(x) \notag \\
    \leq \int \log(2s(x)) \ dF_X(x),
    \label{eq.ch1}
  \end{multline}
  hereby proving the result.
\end{proof}
In the case $\Hyps \{s=t, t\in\R^+_0\}$, one has $H(Y|(X,Y)\in\Tube(X))\leq\log(2t)$.
The motivation for the analysis of the support tube is found in the following 
upper-bound to the mutual information based on a finite sample.
\begin{Theorem}[Lower-bound to the Mutual Information]
  Given an hypothesis class of tubes $\Hyp(\Hypm,\Hyps)$  and a set of i.i.d. samples $\Data$.
  Let $\epsilon(\delta,D,n)$ as in equation (\ref{eq.tolerance}) 
  for a confidence exceeding $1-\delta<1$, and assume that the corresponding 
  probability of covering satisfies $\epsilon(\delta,D,n)<0.5$.
  The following lower bound on the expected mutual information $I(\Y|\X)$ holds
  with probability exceeding $1-\delta$
  \begin{equation}
    H(\Y|\X) \leq \epsilon(\delta,D,n) H(\Y) + (1-\epsilon) \Exp[\log(2s(X))]
    \label{eq.cebound}
  \end{equation}
  and equivalently
  \begin{equation}
    I(\Y|\X) \geq 
    (1-\epsilon(\delta,D,n)) 
    \Big( H(\Y) - \Exp\left[\log(2s(X))\right]\Big) 
      - h(\epsilon(\delta,D,n)),
    \label{eq.mibound}
  \end{equation}
  where $F_{\X}$ denotes the marginal distribution of $\X$ 
  and $h(\cdot)$ is the entropy of a Bernoulli random 
  variable with parameter $\epsilon$.
\end{Theorem}
\begin{proof}
  The proof of this inequality follows roughly the derivation 
  of Fano's inequality as in e.g. \cite{cover91}.
  Let the random variable $\E = g(\X,\Y,\Tube)\in\{0,1\}$ 
  be defined as 
  $\E = \ind(\Y\not\in\Tube(\X))$ with $n$ i.i.d. samples 
  $\left\{\E_i=\ind(Y_i\not\in\Tube(X_i))\right\}_{i=1}^n$.
  Twice the application of the chain rule on the conditional entropy gives
  \begin{eqnarray}
    H(\E,\Y|\X) 
    &=& H(\Y|\X) + H(\E|\X,\Y)= H(\Y|\X)  \\
    H(\Y,\E|\X) 
    &=& H(\E|\X) + H(\Y|\E,\X) \notag \\
    &&\leq H(\E) + H(\Y|\E,\X),
    \label{eq.mutualinformation1}
  \end{eqnarray}
  since $\E$ is a function of $\X$ and $\Y$,
  the conditional entropy $H(\E|\X,\Y)=0$, and $H(\E|\X) \leq H(\E)$.
  Theorem 1 states  that for $\Tube$ with zero empirical risk,
  the actual risk satisfies $\Exp[\E] = \Risk(\Tube;F_{XY}) \leq \epsilon(\delta,D,n)$ 
  with probability higher than $1-\delta$,
  such that the quantity $H(\E)$ can be bounded with the same probability as 
  \begin{equation}
    H(\E) \leq - \epsilon \log(\epsilon) 
    - (1-\epsilon) \log(1-\epsilon) \triangleq h(\epsilon),
    \label{eq.h_epsilon}
  \end{equation}
  because the entropy of a binomial variable is concave with maximum at $0.5$ 
  and $0<\epsilon(\delta,D,n)<0.5$ by assumption, see e.g. \cite{cover91}.

  Now, the second term of the rhs of (\ref{eq.mutualinformation1}) is considered.
  Note first that since $H(\Y) \geq H(\Y|\X,\E=0)$, 
  it holds for all $0<a<\epsilon(\delta,D,n)\leq 0.5$ that 
  \begin{multline}
    a H(\Y) + (1-a) H(\Y|\X,\E=0) \\
    \leq \epsilon H(\Y) + (1-\epsilon(\delta,D,n)) H(\Y|\X,\E=0).
    \label{eq.ineq111}
  \end{multline}
  Hence,
  \begin{gather}
     H(\Y| \E,\X) = P(\E=1) H(\Y|\X,\E=1) \notag \\ 
                  + P(\E=0) H(\Y|\X,\E=0) \notag\\
     \leq P(\E=1) H(\Y) + P(\E=0) H(\Y|\X,\E=0) \\
     \leq \epsilon(\delta,D,n) H(\Y) + (1-\epsilon(\delta,D,n)) H(\Y|\X,\E=0) \notag\\
     \leq \epsilon(\delta,D,n) H(\Y) + (1-\epsilon(\delta,D,n)) \Exp[\log(2s(X))],
     \label{eq.mutualinformation2}
  \end{gather}
  where the first inequality follows from 
  $H(\Y|\X,\E=1) \leq H(\Y)$, and the second one from (\ref{eq.ineq111}) and
  since $P(\E=1)<\epsilon(\delta,D,n)$. 
  The third inequality constitutes the core of the proof, 
  following from the previous Lemma.
  Combining this inequality with (\ref{eq.h_epsilon}) 
  and the definition of mutual information, 
  $I(\Y|\X) = H(\Y) - H(\Y|\X)$  yields inequality (\ref{eq.mibound}).
\end{proof}
In the case of the class of tubes with constant nonzero width $2t\in\R^+_0$,
the inequality can be written as follows. With probability higher than 
$1-\delta<1$, the following lower-bound holds
\begin{equation}
  I(\Y|\X) \geq (1-\epsilon(\delta,D,n)) 
  \left(H(\Y) - \log(2t)\right)  - h(\epsilon(\delta,D,n)),
  \label{eq.mibound2}
\end{equation}
if $\epsilon(\delta,D,n)<0.5$.
Maximizing this lower-bound can be done by 
minimizing the width $t$ and maximizing the probability of covering $(1-\epsilon)$,
since the unconditional entropy is fixed.

From definition 1, it follows that a ST is not uniquely defined for a fixed $F_{\X\Y}$.
From the above derivation, a natural choice is to look 
for the most informative (and hence the least conservative) support tube 
as follows
\begin{equation}
  \Tube^\ast = \argmin_{\Tube\in\Hyp(\Hypm,\Hyps)} \|s\|
  \suchthat \Tube \mbox{\ is a ST to \ } F_{\X\Y}.
  \label{eq.st.leastconservative}
\end{equation}
where $\|\cdot\|$ denotes a (pseudo-) norm on the hypothesis space $\Hyps$,
proportional to the term $\Exp[\log 2s(\X)]$ of equation (\ref{eq.mibound}).
Let the theoretical risk of a ST on $F_{\X\Y}$ be defined as 
$\Risk(\Tube,F_{\X\Y})=\int P\left(\Y\not\in\Tube(x) \ | \ \X=x\right)dF_{\X}$.
Given only a finite number of observations in $\Data$,
the empirical counterpart is studied 
\begin{equation}
  \widehat\Tube = \argmin_{\Tube\in\Hyp(\Hypm,\Hyps)} \|s\|_{\Hypm}
  \suchthat \Risk_n(\Tube;\Data)=0.
  \label{eq.est.leastconservative}
\end{equation}

\subsection{Quantile Tubes}

The discussion can be extended to the case of quantile 
tubes of a level $0<\alpha<1$. 
Assume we have an estimator which for a sample $\Data$ 
returns a tube $\widehat\Tube$ specified 
by exactly $D$ samples such that at most $\lceil\alpha n \rceil$
samples violate the tube. 
The question how well this estimator behaves for novel samples is considered.
Specifically, we bound the expected occurrence of a sample
not contained in the tube $\widehat\Tube$ as follows
using Hoeffding's inequality as classical.
\begin{Proposition}[Deviation Inequality for Quantile Tubes]
  When $\Data$ contains $n$ i.i.d. samples,
  and any hypothesis $\Tube$ can be represented by exactly $D$ 
  samples, one has with probability exceeding $1-\delta<1$, one has
  \begin{equation}
    \Risk(\widehat\Tube;F_{XY}) - \alpha \leq \\
    \Risk_n(\widehat\Tube;\Data) 
    +  
    2\sqrt{\frac{2D\log(\frac{2 n e}{D}) - 2\log\left(\frac{8}{\delta}\right)}{n}}.
  \end{equation}
\end{Proposition}
This proof follows straightforwardly from the Vapnik and Chervonenkis inequality
with $K_{n,D}(\Hyp) \leq \left(\frac{2 n e}{D}\right)^D $ different hypotheses, 
see e.g. \cite{vapnik98} or \cite{devroye96}.
It is a straightforward exercise to use this result to
derive a bound on the mutual information in the case of quantile tubes as previously.

\section{Linear Support/Quantile Vector Tubes}

Given the specified methodology, this section elaborates on
a practical estimator and shows how to extend results 
to quantile tubes. Here we restrict ourselves to the linear model class 
$\Hypm=\{m: m(x)= x^T w \ | \ w\in\R^d \}$ and the class
of parallel tubes $\Hyps=\{s: \ s(x)=t, \ t\in\R^+\}$ 
with constant width for clarity of explanation. 
Problem (\ref{eq.est.leastconservative}) with 
$\Hyp(\R^d,\R^+)$ can be casted as a linear programming problem 
as follows,
\begin{equation}
  (\hat{w},\hat{t}) = \argmin_{w,t>0} \ \ t
  \suchthat -t \leq Y_i - w^TX_i \leq t \ \ \forall i=1,\dots,n.
  \label{eq.est.lintube}
\end{equation}
The more general case of QT requires an additional step:
\begin{Lemma}[Quantile Vector Tubes]
  The following estimator (strictly) excludes 
  at most $C$ observations ({\em quantile property}),
  while the functions $w^T x - t$ and $w^Tx+t$ interpolate at least 
  $d+1$ sample points ({\em interpolation property}).  
  If the underlying distribution $F_{XY}$ is Lebesgue smooth and non-degenerate 
  (hence no linear dependence between the variables and the vector of ones occur),
  exactly $d+1$ points are interpolated with probability 1.
  \begin{multline}
    (\widehat{\mathcal{T}_{w,t}},\xi_i) = 
    \argmin_{w,t,\xi_i} \J_C(t,\xi_i) =  C t + \sum_{i=1}^n \xi_i \\
    \suchthat -t - \xi_i \leq w^T X_i-Y_i \leq t + \xi_i, \xi_i\geq 0
    \ \ \forall i=1,\dots,n.
    \label{eq.qst}
  \end{multline}
  Moreover, the observations which satisfy the inequality constraints exactly 
  determine the solution completely ({\em representer property}), hereby
  justifying the name of Support/Quantile Vector Tubes 
  in analogy with the nomenclature in support vector machines.
\end{Lemma}
\begin{proof}
  The quantile property is proven as follows.
  Let $\alpha_i^+,\alpha_i^-\in\R^+$ be positive Lagrange
  multipliers $\forall i=1,\dots,n$. The Lagrangian of the constrained 
  problem (\ref{eq.qst}) becomes
  $\Lag_C(w,t,\xi_i; \alpha^+,\alpha^-,\beta)$
  $=\J_C(w,t,\xi_i)     - \sum_{i=1}^n \beta_i\xi_i$
  $- \sum_{i=1}^n \alpha_i^+\left( w^T X_i - Y_i + t + \xi_i \right)$
  $- \sum_{i=1}^n \alpha_i^-\left( Y_i - w^T X_i + t + \xi_i \right)$.  
  The first order conditions for optimality become
  \begin{equation}
    \begin{cases}
      \dfrac{\partial \Lag_C}{\partial t} = 0 \rightarrow
      C = \sum_{i=1}^n (\alpha_i^+ + \alpha_i^- ) & (a) \\  
      \dfrac{\partial \Lag_C}{\partial w} = 0 \rightarrow
      0_n = \sum_{i=1}^n \left( \alpha_i^- - \alpha_i^+\right) X_i & (b) \\
      \dfrac{\partial \Lag_C}{\partial \xi_i} = 0 \rightarrow
      1 = (\alpha_i^+ + \alpha_i^-) + \beta_i. & (c)\\
    \end{cases}
    \label{eq.kkt}
  \end{equation}
  Following the complementary slackness conditions 
  ($\beta_i \xi_i = 0 \ \forall i=1,\dots,n$),
  if follows that $\beta_i=0$ for data-points outside the tube 
  ($\xi_i>0$). This together with condition 
  (\ref{eq.kkt}.a) and   (\ref{eq.kkt}.c) proofs the quantile property. 

  The interpolation property follows from the fundamental lemma of a linear programming problem:
  the solution to the problem satisfies at least $d+1+n$ inequality constraints with equality.
  If $\hat{t}\neq 0$, then at least $d+1$ constraints $\xi_i=0$ should be satisfied
  as at most $n$ constraints of the $2n$ inequalities of the form 
  $-t - \xi_i \leq (w^T X_i-Y_i)$ and $(w^T X_i-Y_i)\leq t + \xi_i$ can hold at the same time. 
  If $\hat{t}=0$, the problem reduces to the classical least absolute deviation 
  estimator, possessing the above property.
  Let $x=(X_1,\dots,X_n)^T\in\R^{n\times d}$ be a matrix 
  and  $y=(Y_1,\dots,Y_n)^T\in\R^{n}$ be a vector. 
  If the matrix $(1_N, x, y)\in\R^{n\times (1+d+1)}$ is nonsingular ($F_{XY}$ is non-degenerate) 
  the solution to the problem 
  (\ref{eq.qst}) satisfies exactly $n+d+1$ inequalities, and any two functions 
  $\{w^T x-t, \ w^T x+t\}$ can at most (geometrically) interpolate $d+1$ linear independent points. 

  Since a solution interpolates $d+1$ (linear independent) 
  points exactly under the above conditions, 
  knowledge of which points - say $\mathcal{S}\subset\{1,\dots,n\}$ - 
  implies the optimal solution $\hat{w}$ and $\hat{t}$ as 
  \begin{equation}
    w^T X_i \pm t = Y_i, \ \ 
    \forall i\in\mathcal{S},
    \label{eq.hatw}
  \end{equation}
  where $\pm t$ denotes whether the specific sample interpolates the upper- or lower function.
  This means that the solution can be represented as the set
  $\mathcal{S}$ together with a one-bit flag indicating the sign.
  To represent the solution, one as such needs $(d+1)(\ln(n)+1)$ bits.
  The probability mass inside the tube is given by the value $C$ which is known a priori.
\end{proof}
Note that a similar principle lies at the heart of the derivation of the 
$\nu$-SVM \cite{schoelkopf02}. The representer property is 
unlike the classical representer theorems for kernel machines, 
as no regularization term (e.g. $\|w\|$) occurs in the estimator.
In the case of $C\rightarrow 0$, the estimator 
(\ref{eq.qst}) results in the smallest support tube.
When $C\rightarrow +\infty$, the robust $L_1$ norm is obtained
\cite{rousseeuw86}, 
and when $C$ is such that $t=\epsilon$, the $\epsilon$-loss of the SVR is implemented.
One has to keep in mind however that despite those computational analogies,
the scope of interval estimation differentiates substantially from the $L_1$ 
and the SVR point predictors.

We now turn to the computationally more challenging task of 
estimating multiple condition quantile intervals at the same time. 
\begin{Proposition}[Multi-Quantile Vector Tubes]
  Consider the set of tubes defined as 
  \begin{equation}
    \Tube^{(m)} = \left\{\Tube^l = 
      \left[w^Tx-\sum_{k=1}^l t_k^-, \ \ \ w^Tx + \sum_{k=1}^l t_k^+\right] \right\}_{l=1}^{m}
    \label{eq.setoftubes}
  \end{equation}
  where $m(x) = w^T x$. The parameters $w\in\R^d$, 
  $t^+ = (t_0,\dots,t_m)^T\in\R^{m+1}$ and   $t^- = (t_0^-,\dots,t_m^-)^T\in\R^{m+1}$
  can be found by solving the following convex programming (LP) problem
  \begin{multline}
    \min_{w,t^+,t^-,\xi_i^m} \J_C(t^+,t^-,\xi_i^m) =  
    \sum_{l=1}^{m} C_l (t^+_l + t_l^-) + \sum_{l=1}^m\sum_{i=1}^n (\xi_i^{l+} + \xi_i^{l-}) \\
    \suchthat 
    \begin{cases}
      \vspace{+2mm}
      -\xi_i^{-l} - t_l^-  
      \leq (w^T X_i - Y_i) \leq 
      t_l^+ + \xi_i^{+l} \\
      0 \leq \xi_i^{l+},\xi_i^{l-}, \ \ 
      0 \leq t_l^+, t_l^- 
   \end{cases}\\
   \forall l = 1,\dots,m, \ \ 
   \forall i=1,\dots,n.
   \label{eq.qst}
  \end{multline}
  Then every solution excludes at most 
  $C_l$ datapoints ({\em generalized quantile property}), 
  while the boundaries of all tubes pass through at most $d+2(m+1)$ datapoints.
\end{Proposition}
\begin{proof}
  The proof follows exactly the same lines as in Proposition 5,
  employing the fundamental theorem of linear programming and the 
  first order conditions of optimality. Note that by construction, the 
  different quantiles are properly nested, i.e. not allowed to cross.
\end{proof}

\begin{figure}
  \begin{center}
    \scalebox{.5}{\psfig{figure=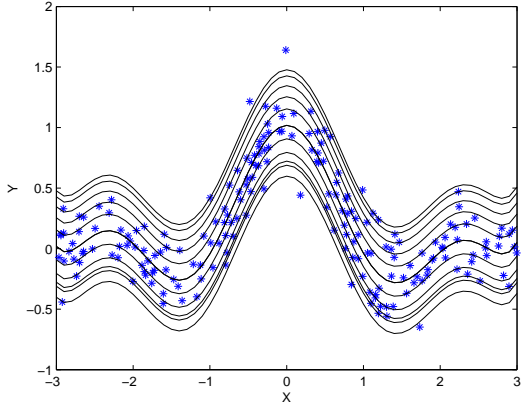}}
  \end{center}
  \caption{\sl\small 
    Example of $n=250$ a Multi-Quantile Vector Tube $\Tube^{(6)}$ with 
    $\alpha = (25,   12, 6,  3, 2, 1)$. Here $m$ consists of 
    a linear combination of 10 localized basis-functions.}
  \label{fig.cqvt}
\end{figure}

Figure \ref{fig.cqvt} gives an example of such a multi-quantile tube 
with a nonlinear function $m$ which is  a linear combination of localized basis-functions.
This computational mechanism of inferring and representing the empirically 
optimal tube $\widehat\Tube$ can be extended to data represented 
in a more complex metric (e.g. $\X\subset\R^{d}$ where $d\rightarrow\infty$,
or by using reproducing kernels).
Hereto, it is easily seen that one needs another mechanism of restricting 
the hypothesis space $\Hypm$. Consider for example the class 
$\Hypm_\rho = \{m(x) = w^T x \ | \|w\|_2^2 \leq \rho\}$,
having a finite covering number (see e.g. \cite{vapnik98}).
The disadvantage in this case is on the one hand 
that one should should choose the regularization constant in an appropriate way {\em a priori}.
On the other hand, the influence of the regularization term becomes nontrivial 
in both the theoretical as well as in the computational derivation.

\section{Conclusion}

This paper \footnote{ {\bf Acknowledgments}

  Research supported by BOF PDM/05/161,
  FWO grant V 4.090.05N, IPSI Fraunhofer FgS, Darmstadt, Germany.
  (Research Council KUL): GOA AMBioRICS, CoE EF/05/006 Optimization in
  Engineering, several PhD/postdoc \& fellow grants; (Flemish
   Government):
    (FWO): PhD/postdoc grants, projects,
        G.0407.02, 
        G.0197.02, 
        G.0141.03, 
        G.0491.03, 
        G.0120.03, 
        G.0452.04, 
        G.0499.04, 
        G.0211.05, 
        G.0226.06, 
        G.0321.06, 
        G.0553.06, 
        G.0302.07. 
        research communities (ICCoS, ANMMM, MLDM);
    (IWT): PhD Grants,GBOU (McKnow), Eureka-Flite2
    - Belgian Federal Science Policy Office:
    IUAP P5/22,
    PODO-II,
    - EU: FP5-Quprodis; ERNSI;
    - Contract Research/agreements: ISMC/IPCOS, Data4s, TML, Elia, LMS, Mastercard.
    JS is a professor and BDM is a full professor at K.U.Leuven Belgium.
}
 studied an intuitive estimator of the conditional 
support and quantiles of a distribution.
The result is shown to be useful to estimate the mutual 
information of the sample by extending the reach of Fano's theorem 
in combination with standard results of learning theory.
It is indicated how the theoretical results relate to 
estimating the minimal convex hull.


\bibliographystyle{IEEEtran}
\bibliography{../../refs}

\end{document}